**Title** Symmetry-protected Spoof Localized Surface Plasmonic Skyrmion

*Jie Yang[1], Xuezhi Zheng[1]\*, Jiafu Wang[2]\*, Yueting Pan[3], Anxue Zhang[2], Tiejun Cui[4]\* and Guy A E Vandenbosch[1]*

Jie Yang, Dr. X. Zheng, Prof. G. A. E. Vandenbosch
WaveCoRE research group, KU Leuven, Kasteelpark Arenberg 10, BUS 2444, Leuven B-3001, Belgium
E-mail: xuezhi.zheng@esat.kuleuven.be

Prof. Jiafu Wang, Prof. Anxue Zhang
Xi'an Jiaotong University, Xianning West Road 28, Xi'an City, 710049, China.
E-mail: wangjiafu1981@126.com

Dr. Yueting Pan
Beijing Computational Science Research Center, Beijing 100193, China.

Prof. Tiejun Cui
State Key Laboratory of Millimeter Wave, Southeast University, Nanjing 210096, China
Email: tjcui@seu.edu.cn

Abstract: Electromagnetic (EM) skyrmions are an EM analogue of the skyrmions in condensed matter physics, which offer new degrees of freedom to structure light and manipulate light–matter interactions and thus promise various groundbreaking applications in optics and photonics. Recently, there is a growing interest in composing EM skyrmions based on different field vectors of EM waves. Here, we realize an EM skyrmion, i.e., a spoof plasmonic skyrmion, using the electric field vectors of spoof localized surface plasmons (spoof LSPs) in a planar microwave resonator with rotational and mirroring symmetries. We construct the spoof plasmonic skyrmion, which holds a hedgehog-like configuration in its electric field vectors, by synthesizing a scalar vortex with a topological charge 0 in the out-of-plane component of the fields, and a polarization vortex with a topological charge 1 in the in-plane component of the fields. Besides an experimental demonstration of this skyrmion, we employ group representation theory and pinpoint the symmetry origin of the skyrmion. Such an investigation demonstrates the ubiquity of the existence of the skyrmion in any planar EM resonator holding rotational and mirroring symmetries, regardless the dimensions and the operating frequencies. The designed skyrmion not only promises novel microwave applications for sensing, processing, storing and transferring information, but also lays down a general guideline for devising skyrmions operating over a broad range in the EM spectra owing to the fact that the conducted symmetry investigation is independent of specific dimension or frequency.

**1. Introduction**

Since firstly proposed by Skyrme to describe the interactions of pions in nuclear physics,[1-2] skyrmions, topologically stable three-component vector field configurations, have been generalized to various subjects in condensed matter physics, including quantum Hall magnets,[3] Bose-Einstein condensates,[4] nematic liquid crystals,[5] etc. Owing to the topological stability even at small sizes, skyrmions promise applications in spintronics and information storage and transfer.[6-8] Motivated by the advances of skyrmions in condensed matter physics and largely fueled by the explosive development of singular optics and topological photonics, optical and plasmonic analogues to skyrmions, i.e., using electromagnetic (EM) fields to emulate the vector configurations of skyrmions, have been attracting growing attention and rapidly have become a cutting-edge topic in optics and photonics. Optical skyrmions were first generated by interfering multiple surface plasmon polariton waves.[9] Subsequently, many approaches have been proposed to construct optical skyrmions using different field vectors, such as the electric or magnetic field vectors of free-space or confined waves,[10-13] spin vectors of the evanescent EM waves,[14-16] synthesized Stokes vectors of paraxial vector beams,[17-18], and synthesized pseudospin vectors in photonic crystals,[19] etc. These optical skyrmions exhibit great topologically protected stability and deep-subwavelength features, enable new degrees of freedom to engineer structured light and light–matter interactions, and promise many applications in nanometer- and femtosecond-scale metrology,[16,20] deep-subwavelength microscopy,[14] ultrafast vector imaging,[12] and topological Hall devices.[19]

Very recently, a new method to construct optical skyrmions was reported based on magnetic field vectors of magnetic localized surface plasmons (LSPs).[21] This work opens the avenue for the fundamental and applied explorations of skyrmions at lower frequencies, where the EM skyrmions can be generated and observed in a simpler manner compared with the optical band.[21] EM skyrmions at microwaves have shown potential for use in many advanced microwave applications like ultra-compact and topologically stable plasmonic devices, including but not limited to ultra-small antennas, wearable microwave sensors, and so on. However, this work only explored the generation of the EM skyrmions based on the magnetic field vectors of the magnetic LSP. Research on EM skyrmions constructed by the electric field vectors of spoof LSPs, which is the electric counterpart of magnetic LSPs,[22-24] has remained untapped.

Here, we propose a design for generating EM skyrmions using the electric field vectors of spoof LSPs in a planar microwave plasmonic resonator with rotational and mirroring symmetries. We numerically and experimentally demonstrate that a spoof plasmonic skyrmion with the skyrmion number of 1 is supported by the resonator. Such a skyrmion is generated by synthesizing a trivial scalar vortex with a topological charge 0 in the out-of-plane component of the electric fields and a polarization vortex with a topological charge 1 in the in-plane component of the electric fields. Besides, by tracing the symmetry origins of the scalar and the polarization vortices with group representation theory, we explore the symmetry origin of the spoof plasmonic skyrmion. In details, we demonstrate that the hedgehog-like configuration of the skyrmion is intrinsically associated with the "identity" irreducible representation (irrep) of the group formed by the symmetries of the resonator.[25] Since the identity irrep always exists for a group formed by rotational and mirroring symmetries, we can conclude that the skyrmion ubiquitously exists in any planar EM resonator with the required symmetries. This conclusion is independent from the dimensions and the operational frequencies of the EM resonator. Therefore, with appropriate excitations, the skyrmion can be observed in generic EM systems with rotational and mirroring symmetries, such as microring resonators,[26-27] circular nanopillars,[28-29] plasmonic vortex lenses,[30-32] photonic quasicrystals,[33] circular nanoemmiter arrays,[34-35] and circular antenna arrays.[36]

## 2. Realization of spoof LSP skyrmion.

We design a five-layer device to generate spoof LSP skyrmions, as shown in **Figure 1**. The first layer is a microwave plasmonic resonator which is made of copper film and holds 8-fold rotational symmetries and mirroring symmetries. The resonator is etched on the second layer made of Rogers 4530B dielectric sheet with 0.508 mm thickness (with relative permittivity 3.48 and loss tangent 0.0037). The same dielectric sheet is used for the third and the fifth layer. Two microstrip lines are etched on the top surface of the third layer and the bottom surface of the fifth layer, respectively, as shown in Figure 1b-c. A via is used to connect the two microstrip lines (via 1 in Figure 1b-c). The fourth layer is made of copper film which serves as the ground of the two microstrip lines. A circular opening is cut in the fourth layer to avoid an electric connection between via 1 and the ground, as shown in Figure 1c. The third to fifth layers contain a feeding network to excite the resonator. The incident waves are fed to the feeding network by a SMA connector from the position marked by "port" in Figure 1a and b. The selection of the feeding topology is closely related with the symmetries of the resonator. This will be postponed to the later part of this work where the symmetry origin of the skyrmion is discussed.

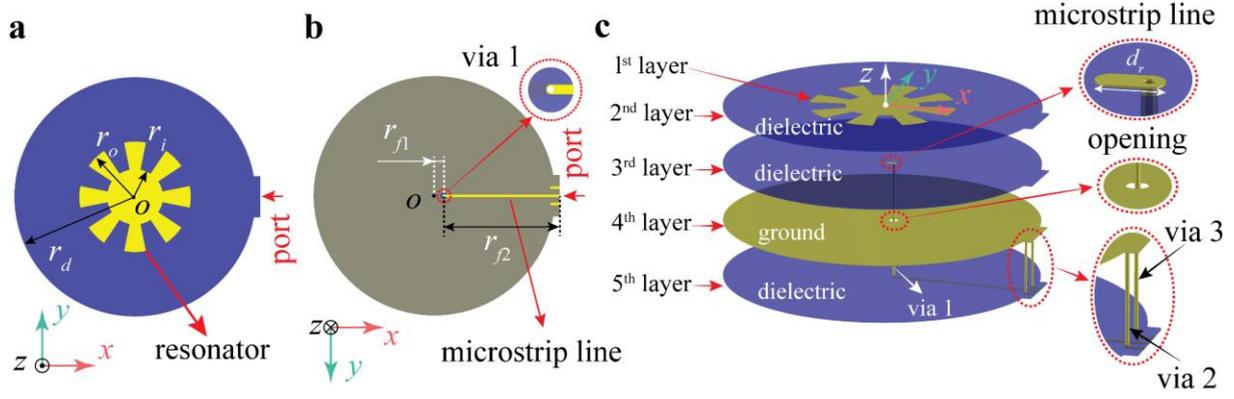

**Figure 1**. Device proposed to generate a spoof LSP skyrmion: (a) top view, (b) bottom view, (c) stratified layers. The adopted parameters are: inner radius of the resonator $r_i = 12$mm, outer radius of the resonator $r_o = 24$mm, radius of the circular dielectric sheet $r_d = 52$mm, angular size of one part of the resonator is $\pi/4$, the distance between the center of via 1 and the center $r_{f1} = 4$mm, the distance between the center of via 1 and the port (i.e., the length of the microstrip line on the bottom surface of the fifth layer) $r_{f2} = 50$mm, and the length of the microstrip line on top of the third layer $d_r = 4$mm. The width of the two microstrip lines is 1.1mm yielding the input impedance of 50 Ω, matched with the SMA connector. In (c), two additional vias marked by "via 2" and "via 3" are added to connect the ground of the SMA connector with the ground of the microstrip lines (i.e., the fourth layer). The radius of the opening is 1.5mm.

The whole device was designed in CST microwave studio. Corresponding simulation results are shown in **Figure 2** and partially in **Figure 3**, in which Figure 2a depicts the magnitude distribution of the electric field and the configuration of the normalized electric field vectors. From Figure 2a, it can be observed that the electric field vector is "up" at the center where $r = 0$, gradually flips as the radius increases, and finally points "down" at $r = R_s$. This electric field configuration defines an "electric field-based skyrmion".[9,14]

Further, we evaluate the topological invariant of the "electric field-based skyrmion", i.e., the skyrmion number.[9] The skyrmion number describes, when the vector distribution is mapped onto the unit sphere, how many times the vectors cover a unit sphere. The skyrmion number can be evaluated as below.[9]

$$s = \frac{1}{4\pi} \iint_A \mathbf{e} \cdot \left( \frac{\partial \mathbf{e}}{\partial x} \times \frac{\partial \mathbf{e}}{\partial y} \right) dxdy. \tag{1}$$

In the above equation, $\mathbf{e}$ denotes the normalized electric field vector, i.e., $\mathbf{e} = \text{Re}\{E_x, E_y, E_z\}/|\mathbf{E}|$, where $E_{x,y,z} = |E_{x,y,z}|e^{-i\omega t}$, and $A$ is the integration area (the circular region with radius $R_s$ in our case, see Figure 2a). Each normalized electric field vector in Figure 2a can be expressed with its longitude and latitude angles $(\alpha, \beta)$ (see illustration of $\alpha$ and $\beta$ in **Figure S1**),[13,17,20] i.e.,

$$\mathbf{e}(x, y, z) = \left( \cos\alpha(\varphi)\sin\beta(r), \sin\alpha(\varphi)\sin\beta(r), \cos\beta(r) \right), \tag{2}$$

where $r$ and $\varphi$ denote the radial distance and azimuth in a cylindrical coordinate system, respectively. Due to the axial symmetry of the electric field distribution, $\alpha$ and $\beta$ are functions of $\varphi$ and $r$, respectively. Then, the skyrmion number can be evaluated in a closed form (see derivations in SI 1),

$$s = \left( -\frac{1}{2}\cos\beta(r) \Big|_{r=0}^{r=R_s} \right) \left( \frac{1}{2\pi} \int_0^{2\pi} \frac{d\alpha(\varphi)}{d\varphi} d\varphi \right). \tag{3}$$

Equation 3 shows that the skyrmion number is a product of two independent factors. The first factor is related with the number of flips of the unit vectors along the radial direction. The unit vector is "up" at the center $r = 0$, i.e., $\beta = 0$ and $\cos\beta = 1$, and flips "down" at the boundary ($r = R_s$), i.e., $\beta = \pi$ and $\cos\beta = -1$, as shown in Figure 2b and c. Consequently, the first factor is equal to 1. The second factor is related with the variation of the unit vectors along the azimuthal direction. It is observed that along this direction, the relative orientation of the electric field vectors is invariant with respect to the radial direction, i.e., the axial symmetry mentioned above. In other words, the longitudinal angle $\alpha(\varphi)$ is constant over the azimuthal direction and, as a result, $d\alpha/d\varphi=1$. Therefore, we obtain a skyrmion number of the generated spoof LSP skyrmion of 1.

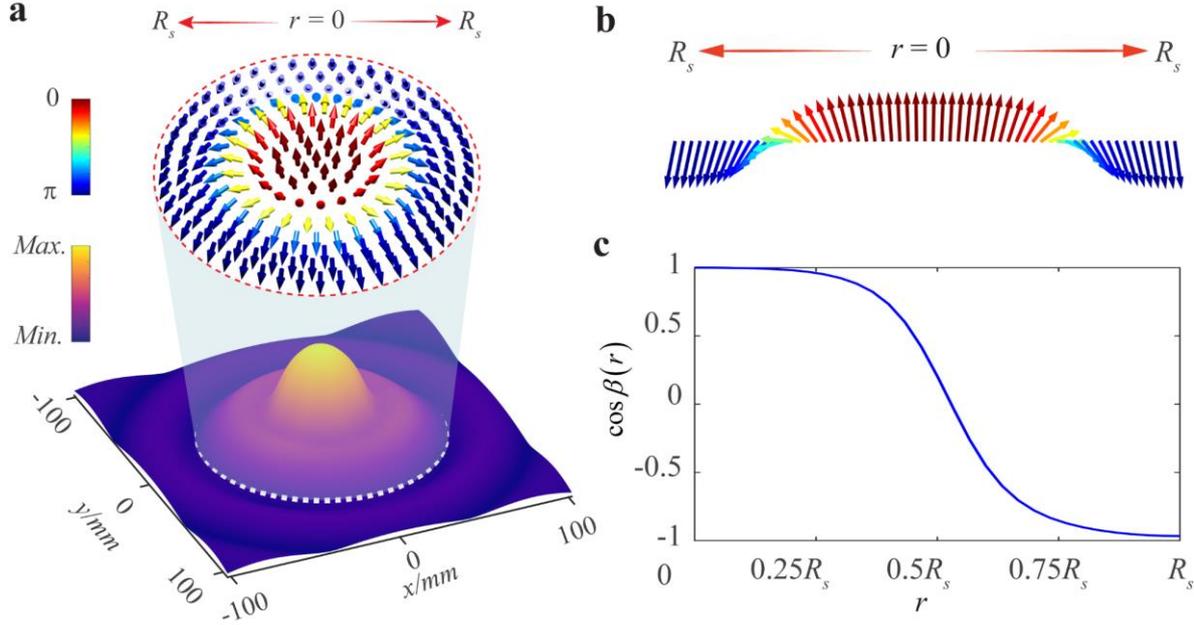

**Figure 2**. Illustration of the generated spoof LSP skyrmion. The electric field distribution at 3.77GHz in the plane 10 mm above the resonator ($z = 10$mm) is chosen to represent the vector configuration of the generated skyrmion. The frequency is obtained from the simulated reflection coefficient $S_{11}$ of the device, as shown in Figure 3. In (a), the bottom and top insets show the magnitude distributions of the electric field (i.e. $|\mathbf{E}| = \sqrt{E_x^2 + E_y^2 + E_z^2}$) and the normalized electric field vectors in the central region. The central region is defined by a white dashed circle in the magnitude distribution. (b) shows the distribution of the normalized electric field vectors along the radial direction. (c) plots the cosine of the latitude angles $\beta$ of the normalized electric field vectors. In (a) and (b), $r$ is the radial distance, and $R_s = 55mm$. The colors of the vectors in (a) and (b) are coded from blue to red to denote the latitude angle $\beta$ varying from $\pi$ to 0.

To verify the above numerical results, the five-layer device was fabricated (see the insets of Figure 3a), and measured. It can be observed that the measured reflection coefficient $S_{11}$ (see Figure 3a) agrees well with the simulated one. The spoof LSP skyrmion is formed at the resonant dip $M0$, which is located at 3.77GHz in the simulation and at 3.72GHz in the measurement. This small discrepancy might arise from numerical or fabrication errors, e.g., in the fabrication glue is used to bond the resonator and the feeding system together, which may lead to the discrepancy. Further, the $z$ component of the electric field over an area of 80×80 mm² at $z = 10$ mm above the device at 3.72GHz is measured (see SI 5 for the measurement setup). The magnitude and phase distributions of the measured $E_z$ are shown in Figure 3d and e. As a comparison, the magnitude and phase distributions of the simulated $E_z$ at $z = 10$ mm at 3.77GHz are shown in Figure 3b and c. It can be observed that the experimental and simulation results agree well.

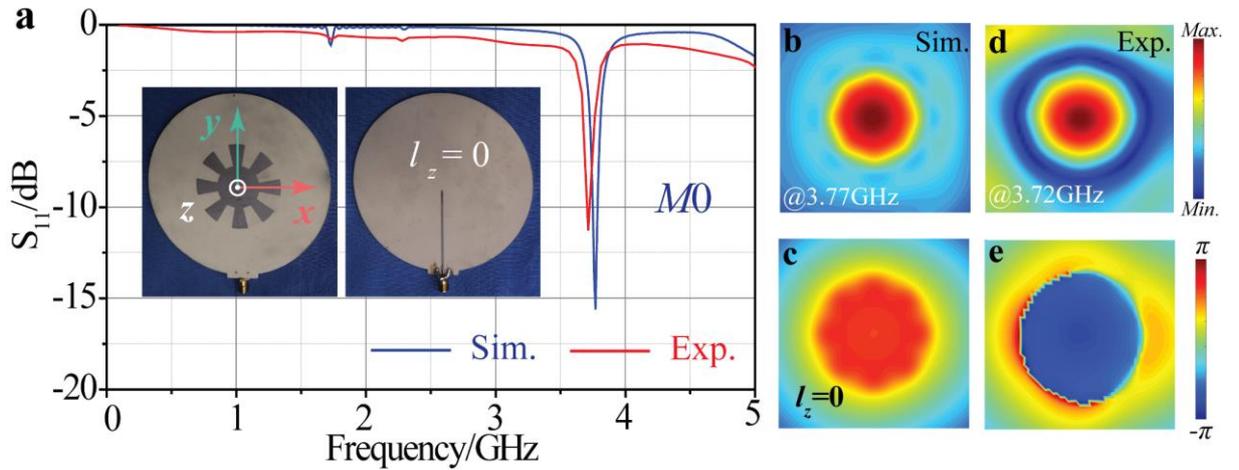

**Figure 3**. Comparison of experimental and simulation results: (a) reflection coefficient $S_{11}$, (b) magnitude and (c) phase distributions of the simulated $E_z$, and (d) magnitude and (e) phase distributions of the measured $E_z$. In (a), the top and bottom views of the fabricated sample are shown. Both the simulated and measured $E_z$ show a trivial scalar vortex mode with topological charge of 0, i.e., $l_z = 0$.

## 3. Symmetry Origins of the Skyrmion.

Firstly, we look for the topological signatures of the electric field distributions underlying the skyrmion excited at 3.77 GHz. On the one hand, it can be observed from Figure 3b and 3c that the out-of-plane component, i.e., the $z$ component, of the electric field always exhibits a trivial scalar vortex mode with the topological charge of 0, i.e., a vortex mode carrying an orbital angular momentum (OAM) of 0.[37] On the other hand, it can be observed from **Figure 4a** and 4b that the in-plane components, i.e., the $x$ and the $y$ components, of the electric field demonstrate a zero magnitude at the center and are linearly polarized over the region of interest. Both aspects indicate that the in-plane components of the electric field form a polarization vortex with a V-type polarization singularity, i.e., a V point. At this singularity, by definition, the magnitude is zero and the polarization azimuth is undefined.[38] By evaluating the Stokes fields (see SI 2), the topological charge of this polarization singularity is found to be 1. Further, it is well-known that a polarization vortex can be conveniently understood in terms of the left and right circular bases, i.e., $\hat{L} = \frac{1}{\sqrt{2}}(\hat{x} + i\hat{y})$ and $\hat{R} = \frac{1}{\sqrt{2}}(\hat{x} - i\hat{y})$, and can be seen as the superposition of the scalar vortex modes for the left and the right circular components,[38-39] i.e., $E_L$ and $E_R$. If the scalar vortex modes corresponding to the left and right circular components have topological charges $l_L$ and $l_R$, the topological charge of the polarization vortex $I$ can be proven to be $\frac{1}{2}(l_R - l_L)$. By projecting the in-plane components of the electric field in Figure 4a and 4b onto the circular bases, it is found that the left- and the right- circular components hold topological charges $-1$ and $+1$. Therefore, the excited skyrmion is composed of three scalar vortex modes (in the $E_L$, $E_R$ and $E_z$ components of the electric field, respectively), and the topological charges of these three scalar vortex modes are $-1$, $+1$ and $0$, respectively.

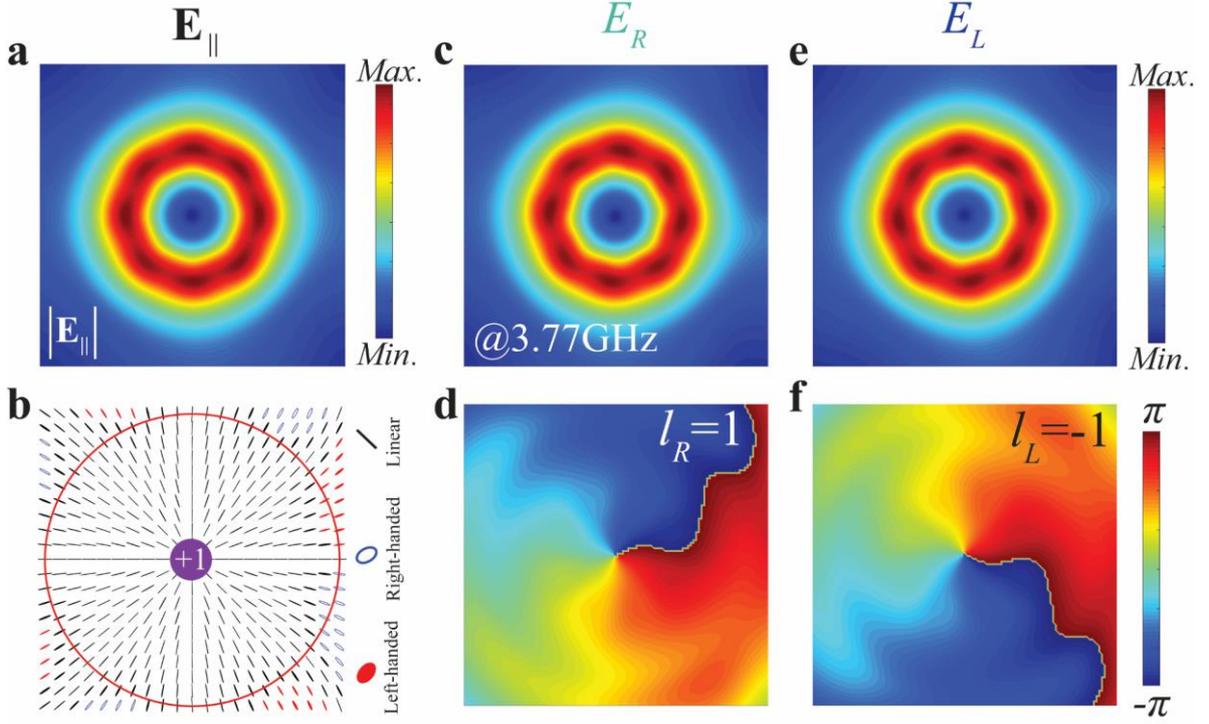

**Figure 4**. Illustration of the in-plane electric field $\mathbf{E}_\parallel$ at 3.77GHz. (a) and (b) show the magnitude and local-polarization-state distributions of the in-plane electric field $\mathbf{E}_\parallel$, respectively. The transverse electric field can be expressed as $\mathbf{E}_\parallel = E_x\hat{x} + E_y\hat{y} = E_R\hat{R} + E_L\hat{L}$, and thus $|\mathbf{E}_\parallel| = \sqrt{E_x^2 + E_y^2}$. (c) and (d) show the magnitude and phase distributions of $E_R$, respectively. (e) and (f) shows the magnitude and phase distributions of $E_L$, respectively. $E_R$ and $E_L$ correspond to scalar vortex modes with topological charges $l_R = 1$ and $l_L = -1$, respectively. The radius of the red solid circle in the bottom right-hand panel is $R_s$, which marks the boundary of the domain where the spoof LSP skyrmion appears.

The topological charges of the scalar vortex modes in $E_L$, $E_R$ and $E_z$ are intrinsically related with the symmetries of the resonator. The rotational and the mirroring symmetries form a $D_8$ point group. To simplify the discussions and to stress the symmetry features, we select eight exemplary spatial points, which can be transformed into each other by the symmetry operations forming the $D_8$ point group in the resonator (see **Figure 5a**), and we impose a polarizable dipole at each point. The discussion on the eight symmetry-related points can be readily extended to the whole resonator, by seeing the resonator as a collection of these points and by reconstructing the currents flowing in the resonator by putting the dipoles at these points. According to the group representation theory, the $D_8$ point group has seven irreps, including four one-dimensional and three two-dimensional irreps.[25] As a result, it can be proven that the polarizable dipoles at the eight points can form seven unique orthogonal dipole distributions (see SI 3 for more information). Especially, the distribution corresponding to the "identity" irrep (see Figure 5a) is of special importance. On the one hand, at each point, the orientation of the dipole is invariant with respect to the radial direction (in the case of Figure 5a, the orientation is along the radial direction). On the other hand, the phases of all the dipoles are the same. The former implies that the electric fields radiated by the dipoles have $E_\rho$ and the $E_z$ components, and that the $E_\varphi$ component is expected to be significantly smaller, even almost zero (**see Figure S5**). The latter requires that $E_\rho$ and $E_z$ hold trivial topological charges (see Figure S5), i.e., a constant phase along the azimuthal direction.

Converting to the circular basis, the trivial topological charge in the $E_\rho$ component leads to the $-1$ and $+1$ topological charges in the $E_L$ and the $E_R$ components (see the derivations in SI 3.5), respectively. Considering the $E_L$, $E_R$ and $E_z$ components, the electric field vectors may demonstrate an "electric field-based" skyrmion as a whole (see Figure 5b). Since the electric field is radiated by the dipole distribution corresponding to the "identity" irrep, the radiated skyrmion is said to belong to the "identity" irrep. The above discussion reveals the existence of an "electric field-based" skyrmion in the targeted resonator. Since only symmetries are involved and the "identity" irrep always exists in a generic $D_M$ group (where $M$ is an integer),[25] the ubiquity of the existence of such skyrmion in any planar EM resonator holding the symmetries forming the $D_M$ group is demonstrated.

To excite the "electric field-based" skyrmion belonging to the "identity" irrep, the symmetries of the resonator require an incident field satisfying a so-called symmetry matching condition (see details in SI 3.3). In mathematical terms, the incident field must have a non-vanishing projection along the identity irrep, and in physical terms, the symmetry of the incident field is at least partially aligned with the symmetry of the electric field vectors of the skyrmion. Take our design in Figure 1 as an example. In the device, the microstrip line on the top surface of the third layer is used to excite the resonator. This microstrip line is quasi-static at the resonant frequency (a length of 4 mm is much smaller than the resonant wavelength of around 80 mm at 3.77 GHz) and holds a mirroring symmetry with respect to the $x$ axis (see Figure 1c). As a result, the generated incident field has a mirroring symmetry with respect to the $x$ axis as well. This symmetry of the incident field is partially aligned with the axial symmetry of the electric field vectors of the skyrmion. Hence, the skyrmion is excited in the device. It is worth mentioning that the single port feeding network in our design is used only for demonstrative purposes, in the sense that the feeding network is the simplest design that can both excite the targeted skyrmion and provide an intuitive picture for the symmetry matching condition. This feeding can be readily generalized to more complex topologies, e.g., multiport feeding networks, or, alternatively, more complex incident fields can be used, e.g., hypergeometric-Gaussian beams with zero total angular momentum,[37] so that the incident field is more efficiently coupled to the skyrmion.

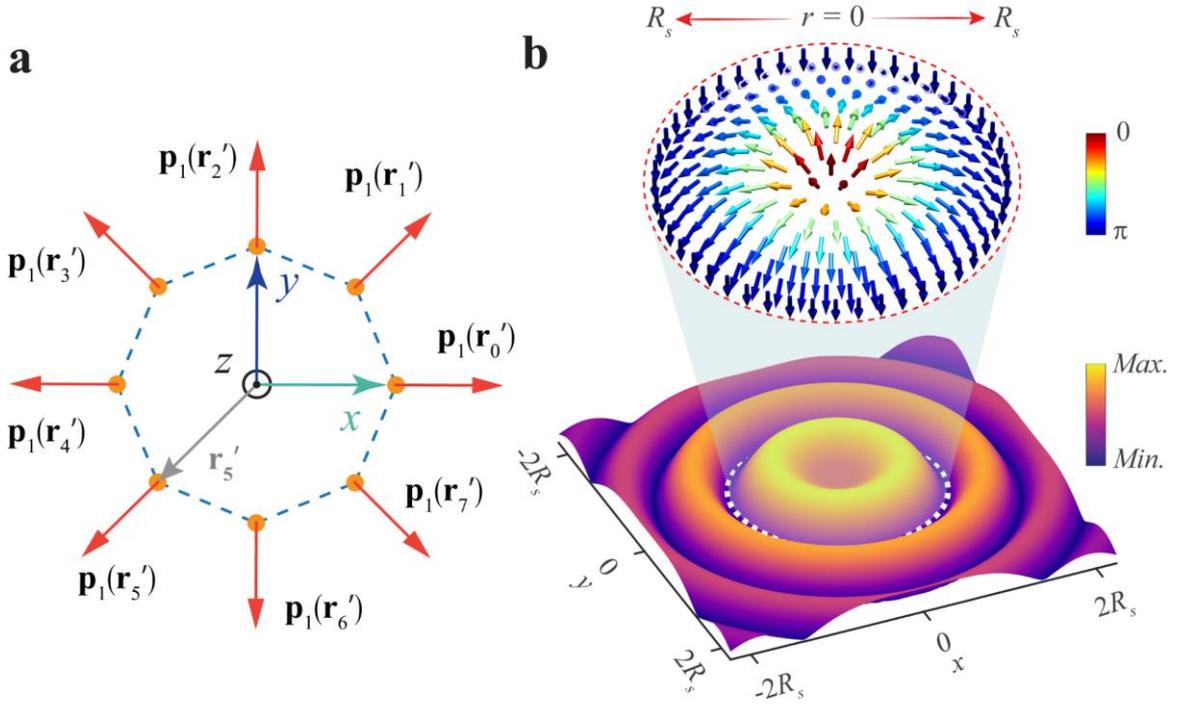

**Figure 5**. The "electric field-based" skyrmion generated by the dipole distribution belonging to the identity irrep of the $D_8$ group: (a) the eigen electric dipole set belonging to the identity irrep; (b) the generated electric-vector-based skyrmion. In (a), the eight dipoles are located in the $xoy$ plane and oscillate along the radial direction at frequency $f_d$ (corresponding to the wavelength $\lambda_d$). The oscillating frequency is chosen as 1GHz in our simulation. It is worth noting that the oscillating frequency of the dipoles actually can be freely chosen because the existence of the skyrmion is independent of the oscillating frequency. Obviously, all dimensions used here are scaled with respect to the oscillating wavelength. The electric dipole momentum of the electric dipole located at $\mathbf{r}_m'$ is marked by $\mathbf{p}_1(\mathbf{r}_m')$, where $m = 0, 1, ..., 7$ (see details in SI 3.4). The distance from the center to each golden point is set as $\lambda_d/8$. In (b), the electric field distribution in the plane $z = \lambda_d$ is chosen to represent the vector configuration of the generated skyrmion, and $R_s = 14\lambda_d/15$. The white dashed circle in (b) marks the boundary of the electric-vector-based skyrmion. The colors of the vectors in (b) are coded from blue to red to denote the latitude angle $\beta$ of the normalized electric field vector varying from $\pi$ to 0.

## 4. Conclusion

We present a design to generate a spoof plasmonic skyrmion by using the electric field vectors of the spoof LSPs. A device consisting of a microwave plasmonic resonator with the $D_8$ group symmetries and a feeding network is designed to launch the skyrmion. Both the simulation and experimental results well demonstrate the formation of the spoof plasmonic skyrmion. By analyzing the spatial configurations of the out-of-plane and in-plane components of the electric fields, we find that a scalar vortex mode with topological charge of 0 and a polarization vortex with topological charge of 1 are formed in the out-of-plane component and in the in-plane components, respectively. These vortex modes constitute the hedgehog-like vector configuration of the spoof plasmonic skyrmion. By exploiting group representation theory, we find that the constituent vortex modes and the skyrmion are intrinsically related with the "identity" irrep of the group formed by the symmetries of the resonator. Since the "identity" irrep of a $D_M$ group always exists, the excited skyrmion is ubiquitous in generic planar EM systems holding the corresponding symmetries.

## 5. Materials and Methods

*Numerical simulations.* Codes are composed in MATLAB to generate the transformation operators, the irreps and the projection operators for the $D_8$ group, and to calculate the electric fields radiated by the eigen dipole set belonging to the identity irrep.

*Experimental measurements.* The fabricated samples are characterized by a near-field scanning system in an anechoic chamber. The scanning system consists of a servo actuator, a coaxial near-field probe, a Vector Network Analyzer (VNA) and connection cables. A 50Ω SMA connector receiving the input signal from the VNA is welded onto the microstrip line. The experiment setup is shown in **Figure S9**. In the fabrication, glue is used to bond the resonator and the feeding system together, which may also lead to discrepancies between simulations and experiments.